# Micromanufacturing of geometrically- and dimensionally-precise molecular single-crystal photonic micro-resonators via focused ion beam milling

*Vuppu Vinay Pradeep and Rajadurai Chandrasekar\**

School of Chemistry and Centre for Nanotechnology

University of Hyderabad, Prof. C. R. Rao, Gachibowli, Hyderabad 500046, India

E.mail: r.chandrasekar@uohyd.ac.in

Highly reproducible manufacturing of organic optical crystals with well-defined geometry and dimension is important to realize industrially relevant all-organic microelectronic and nanophotonic components, and photonic integrated circuits. Here, we demonstrate programmed shape and size alteration of perylene crystal resonators into disk and rectangular geometries using focused-ion beam milling technique. Due to highly reproducible nature of the employed technique, the fabricated smaller sized disk and rectangular crystal resonators displaying shape and size dependent optical modes are suitable for commercial nanophotonic applications.

Frenkel type molecular optical crystals possess many functional properties such as high exciton binding energy (*1*), exciton-polaritons formation (*2*), high refractive index (*3*), fluorescence (*4*), phosphorescence (*5*), optical non-linearity (*6*), chirality (*6,7*), hardness/softness (*8,9*), bandwidth tunability, surface smoothness (*8*), diverse geometries (*9*) and lightweight. The last decade has seen a tremendous progress in utilizing the naturally grown molecular microcrystals as passive and active optical waveguides (*10,11*), resonators (*12*), modulators (*13*), directional couplers (*14*), circuits (*15*), and lasers (*4*). However, the main bottleneck that hinders the direct utility of molecular single-crystals for commercial high-tech applications such as micro-electromechanical systems (*16*), opto-electronic devices (*17*), photonic integrated circuits (PICs) (*18*) is lack of controllability of crystal morphology. In PICs, the shape and size of microcomponents are the two critical factors that determine the photonic device characteristics. PICs are useful for quantum communication and processing, sensing, and spectroscopy, neuromorphic machineries. Till now, the controllability of microscale shape and size suitable for PICs has been successfully achieved to industrial perfection in silicon (*19*) and silicon-derivates (*20*) and group III-V semiconductors (*21*) using electron-beam-lithography and associated etching processes (*22*). As a result, the above-mentioned inorganic materials are widely employed for the commercial-scale manufacturing of microphotonic components such as waveguides, ring-resonators, disk-resonators, gratings and splitters for the construction of monolithic and hybrid PICs. However, monolithic Si-based devices have several drawbacks: indirect bandgap, narrow range visible-near infrared only absorption, high thermo-optical coefficient, centrosymmetry, passive-only light transport, and hardness (*23*). Further, cutting-edge PICs require innovative materials platform providing non-linear optical properties. Thus, hybrid PICs have been manufactured to blend the advantages of many optical materials on a single chip (*24*)

      Realization of molecular crystal PICs (MCPICs) is a next important development step in nanophotonics (*15,25*). Though molecular crystals possess many functional attributes compared to inorganic materials, their geometry and dimension cannot be controlled with microscale precision during their natural growth. The lack of geometrical and size precisions of crystals severely restricts their entry into the PICs market because the latter demands dimensionally uniform PIC elements with high reproducibility. As a result, the desire to employ molecular crystals as optical elements to create all-organic PICs with commercially viable precision manufacturing technique grows day by day. We

developed a mechanophotonics approach (*15,25*) to control the geometrical shape and dimension of optical crystals and integrated them using atomic force microscopy to fabricate several MCPIC components.

Optical resonators are quintessential components in MCPICs. An optical resonator circumnavigates broad-band light via multiple total internal reflections at the crystal-air interface and subsequently produce optical WGMs via constructive optical interference. The frequency of the modes and their separation (free spectral range, FSR = $\lambda_m - \lambda_{m+1}$) depends upon the size and shape of the resonator as FSR ≈ $1/a$, where $a$ is length or diameter. Recently, we reported nearly square-shaped perylene single-crystal optical resonators emitting whispering-gallery-modes (WGMs) which were grown by sublimation method (*26*). Focused ion beam (FIB) milling has been used for etching polymer films (*27*) and inorganic materials (*28*). FIB milling has also been used to convert perylene single crystals into arbitrary shapes (*29*). Creation of well-defined geometrical shapes such as rings, disks, square, hexagons, octahedrons, and spheres are important to produce WGM microresonators. Amongst these geometries, disk-shaped crystal resonators are important photonic elements to process and route the light into clockwise and counterclockwise directions when coupled with other photonic elements. Here, the geometrical precision of the single crystal resonator is crucial to trap light and observe the optical resonances. For example, the rectangular crystal geometry supports Fabry Pérot (FP) optical resonances by reflecting the light back and forth via their mirror-like opposite facets. To the best of our knowledge, micromachining (by either FIB or electron beam lithography) of a naturally grown crystal into well-defined crystal geometry for direct resonator applications has not been done earlier. Therefore, we envisioned exploiting the FIB milling technique to fabricate two geometrically and dimensionally different perylene microresonators. For that, we planned milling of nearly square-shaped perylene single crystal resonators into disk- and rectangular-resonators which can support WGM and FP resonances, respectively useful for MCPICs application.

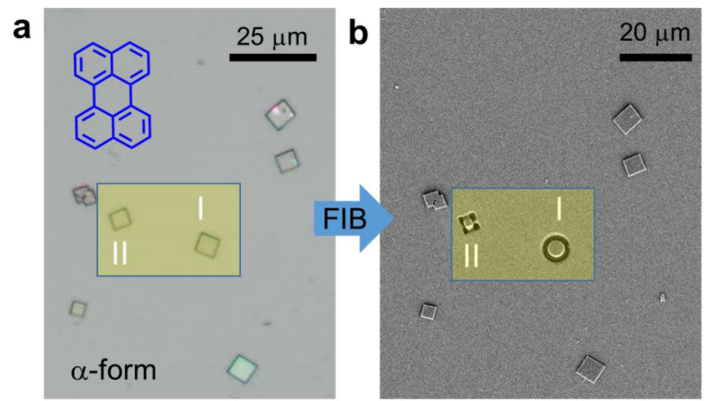

**Figure 1. a.** Confocal optical image of perylene microcrystals (α-form) prepared using bottom-up self-assembly in tetrahydrofuran (4 mM). The yellow area shows the selected crystals I and II for FIB milling. The perylene molecular structure is shown in the inset. **b.** The field emission scanning electron microscopy image (top-view) displaying milled perylene crystals into smaller disk- and rectangular-shapes.

This report demonstrates the first micromachining of perylene single-crystal disk- and rectangular-shaped optical resonators using FIB milling. We used the bottom-up method to naturally grow α-form perylene microcrystals on a clean coverslip (Supporting information Figure 1). Optical microscopy studies reveal the growth of numerous square- and rectangular-shaped microcrystal of various sizes. As test cases, we identified two crystals labelled I and II for the FIB milling (Figure 1a).

As per our plan, the shape of the selected perylene single-crystals I and II were precisely converted to disk- and rectangular-shaped crystals, respectively with size reduction. Remarkably, single-particle micro-spectroscopy studies reveal the micromachined disk- and rectangular-shaped crystals act as microresonators by emitting WGM and FP resonance modes (Figure 1b, 2f, 3f).

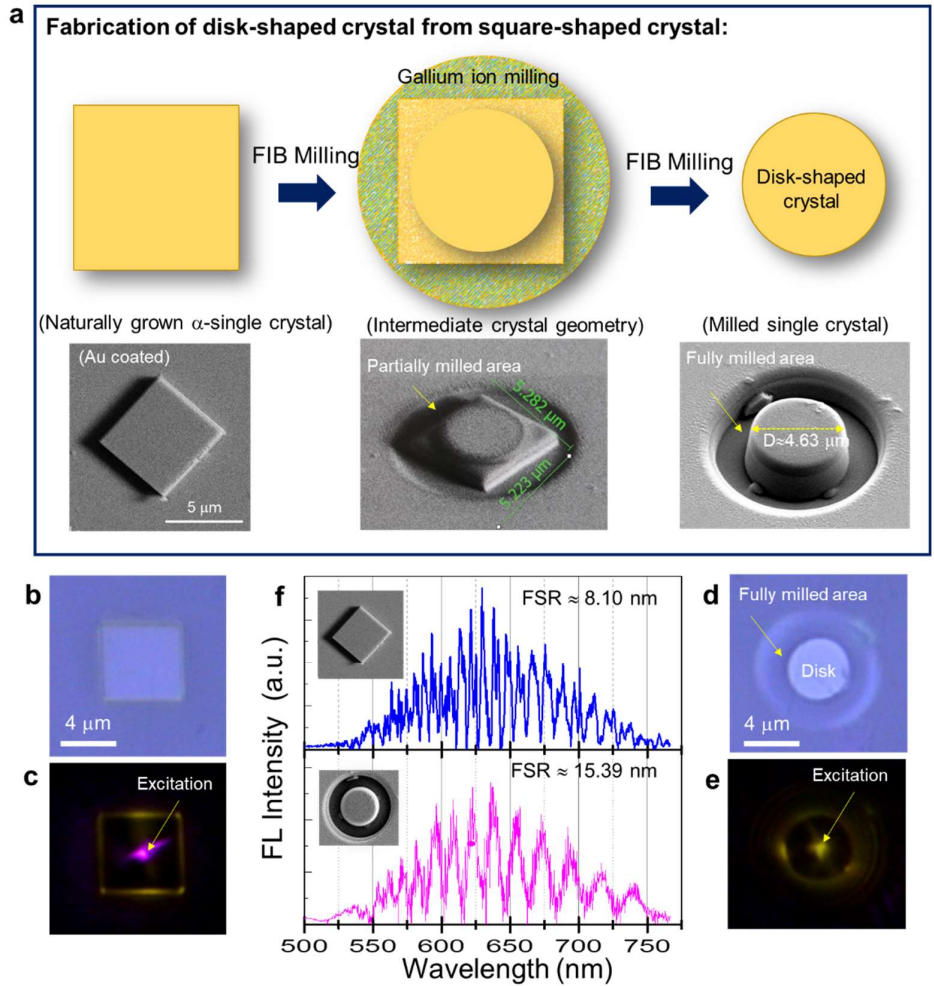

**Figure 2. a.** Schematic of FIB milling of naturally grown square-shaped perylene single crystal into a disk shape. The corresponding FESEM images of crystal before and after milling are shown below the graphics. **b** and **d**. Optical microscopy images of square-shaped perylene single crystal (before milling and gold coating) and a disk-shaped crystal (after milling and removal of gold coating). **c** and **e**. The corresponding fluorescence images. **f.** The fluorescence background-subtracted spectra displaying shape and size-dependent whispering gallery modes.

A nearly square-shaped perylene crystal with ≈5.22× 5.28 µm$^2$ dimensions was identified for micromachining using FIB. Before that, the single-particle micro-spectroscopy experiments were performed in a transmission mode geometry to determine the resonator characteristics of the selected single crystal. When the crystal was excited with a continuous-wave 405 nm laser (Excitation: 0.05 mW; objective: 60x), it displayed a bright yellow fluorescence (FL) at its four edges. The recorded broad FL (objective 150x, numerical aperture: 0.95) spectrum covering the bandwidth of ≈525-775 nm region exhibited a series of pairs (transverse magnetic, TM and transverse electric, TE) of sharp peaks, confirming that the crystal is a WGM microresonator. The FSR value of the crystal resonator is ≈8.10 nm. The resonator characteristics of the square-shaped perylene crystal shown in the field emission scanning electron microscope (FESEM) image arise due to multiple circulations of FL by the four-light-reflective edges of the crystal (Fig. 2a, 3a). Later, the crystal was milled after gold coating using gallium ion beam to fabricate a circular disk-shaped crystal of diameter 4.63 µm (Figure 2a). The gold coating in the sample was removed by repeated washing with KI/Iodine solution. Photonic experiments performed on the disk-shaped crystal exhibited an intense FL spectrum supporting a relatively broad optical modes with an FSR of 15.39 nm. The number of optical modes depends on the geometry and size of the resonators. From the circular shape of the crystal, it is evident that these optical modes occur due to WGM resonances.

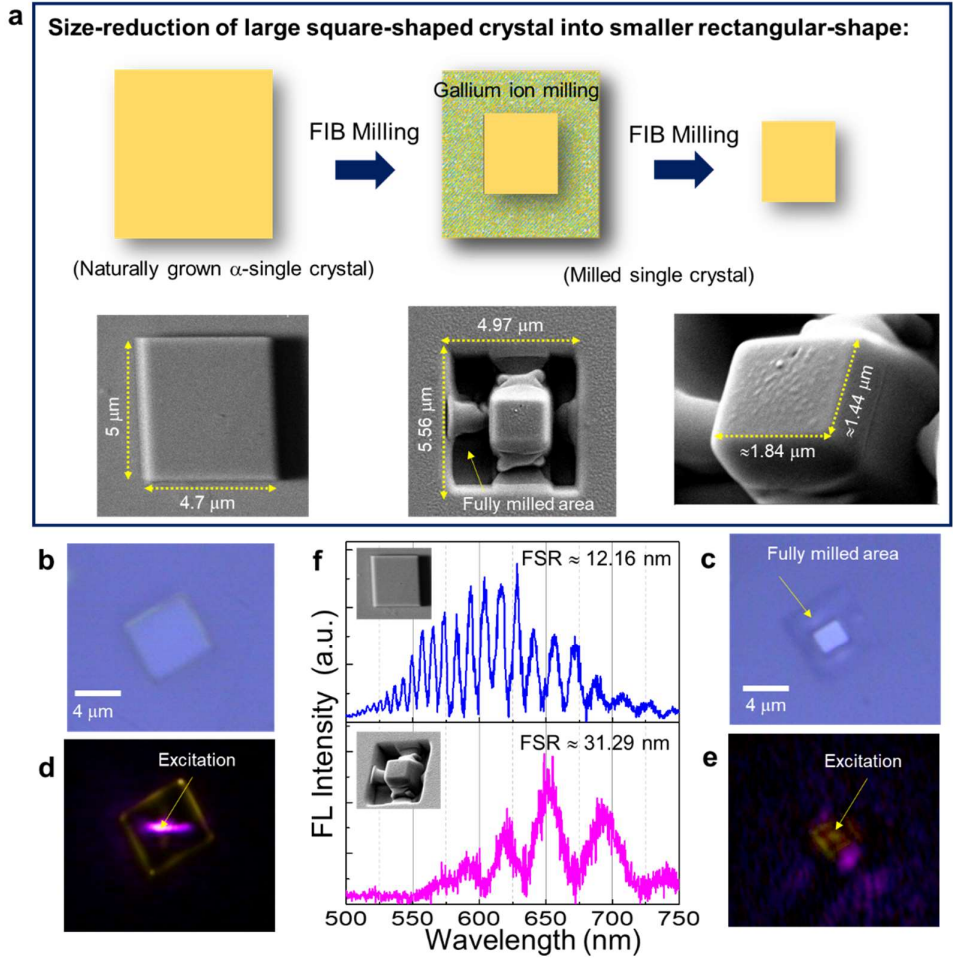

**Figure 3. a.** Schematic of FIB milling of naturally grown rectangular-shaped perylene single crystal into a smaller rectangular shape. The corresponding FESEM images of crystal before and after milling are shown below the graphics. **b** and **c**. Optical microscopy images of rectangular-shaped perylene single crystal (before milling and gold coating) and a smaller rectangular-shaped crystal (after milling and removal of gold coating). **d** and **e**. The corresponding fluorescence images. **f.** The fluorescence background-subtracted spectra displaying size-dependent WGM and FP resonances.

To reduce the size of a large microresonator crystal into a small rectangular microresonator via FIB milling process, a naturally grown perylene crystal of size ≈5×4.7 µm$^2$ was selected. Before milling, the crystal during single-particle micro-spectroscopy experiments exhibited WGM resonances with an FSR value of 12.16 nm. Milling the crystal into a smaller rectangular crystal of dimensions 1.84×1.44 µm$^2$ and subsequent optical experiments display relatively broader modes with FSR of 31.29 nm. The increase in the full-width-at-half-maximum of FP modes and FSR values is in line with the inverse relationship of FSR with the resonator dimension. Further, the crystals were stable to up to 20 mW laser pump power.

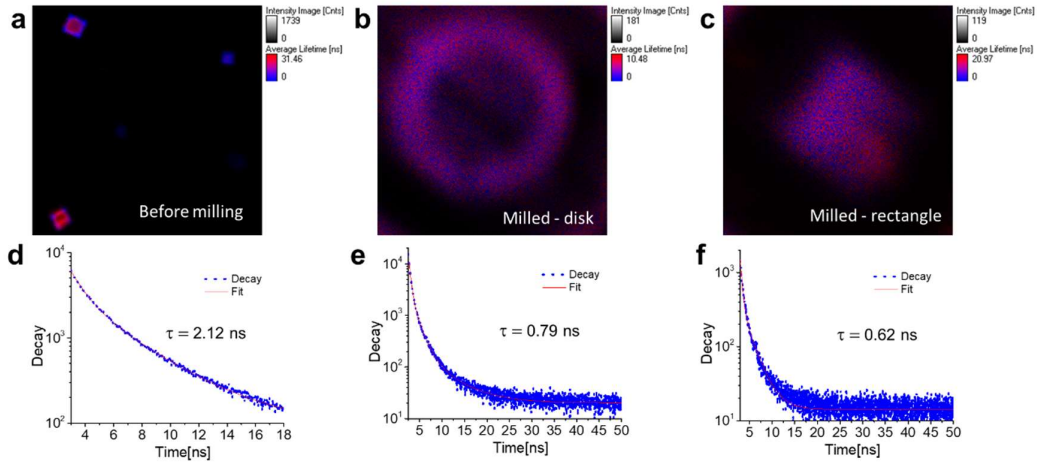

**Figure 4. a-c.** The FL lifetime images of a perylene crystal before and after milling respectively. **d-f.** The corresponding FL lifetime decay plot with estimated average lifetime values.

The FL lifetime of perylene crystals before and after gallium ion milling was investigated using a FL lifetime microscope (pico-second 405 nm pulse laser) with a time correlated single photon counter (Figure 4a-c). The images show distribution of FL lifetime values within the crystal. The disk-shaped resonator showed a well-resolved image with high FL signal from the rim of the circular cavity due to circumnavigating light at the crystal-air interface. On the other hand, for the smaller rectangular cavity, a nearly equal spread of FL was observed. Unlike ordinary crystals, the lifetime values of crystal resonators are different as the quality factor of the resonator determines the photon lifetime (trapped light) of the FL within the crystal by the relation, $Q \approx \tau_P$. The average lifetime decay values of milled crystals are slightly lowered compared to crystals before milling (Figure 4d-f).

In conclusion, FIB milling has been employed to machine perylene crystals resonators into different geometry and size for the first time. The fabrication of disk- and rectangular-shaped photonic resonators is a proof-of-principle experiment that can also be applied to other molecular crystals. The presented technique can be used directly to fabricate circular-, ring-, rod-shaped, and any possible geometries to create photonic modules such as resonators, waveguides, lasers, interferometers, and gratings, couplers, modulators and photonic crystals suitable for the fabrication of PICs. As the geometry and dimension of the molecular crystals can be precisely controlled down to microscale during the milling process, this technique can also be applied to the industrial-scale production of photonic modules for PICs.

# Supporting Information

**Preparation of microcrystals:**

α-form of perylene microcrystals were used in this Example. Bottom-up method was used to naturally grow α-form perylene microcrystals on a clean coverslip (Figure 1). Tetrahydrofuran (HPLC grade) solution of perylene (1 mg/1 mL, 4 mM) was used to fabricate micro-sheets. The solution was sonicated for 30 s and kept for 5 min without any disturbance. Later, 2-3 drops (~20 µl) of perylene solution was drop-casted on the clean glass coverslip and allowed to evaporate. As the solvent evaporates, the micro-sheets started growing. Finally, the solvent got evaporated completely and resulted in micro-sheets of different sizes. The growth of numerous square- and rectangular-shaped microcrystal of various sizes were observed and confirmed by an optical microscopy study.

**FIB milling:**

For the milling of micro-sheets, Thermo Scientific SCIOS 2 Dual Beam instrument was used. Initially, the micro-sheets were imaged in SEM mode with a 0.0° tilt angle, 0.4 nA beam current and an accelerating beam voltage of 5.00 kV. Later, the sample was tilted 52° to align orthogonal to FIB (Gallium ion source). Using a pre-defined shapes in the software (in this case rectangular and circular shapes), the milling portion was selected. Then, 30 kV accelerating beam voltage was applied to mill the micro-sheets to desired shapes.

a) To fabricate a circular disk-shaped crystal of diameter 4.63 µm (Figure 2a), the crystal (I) was milled after gold coating using gallium ion beam. The gold coating in the sample was removed by repeated washing with Lugol's Iodine solution. The energy dispersive X-ray analysis (EDAX) indicated the presence of only a tiny fraction of Ga ions on the sample.
b) To fabricate a rectangular-shaped crystal of dimension 1.84×1.44 µm² (Figure 3a), the crystal (II) was milled after gold coating using gallium ion beam. The gold coating in the sample was removed by repeated washing with Lugol's Iodine solution.

**Gold coating:**

The coverslip containing micro-sheets was placed in QUORM Tech. sputter coater. By applying 15 µA current for about 80 sec, a thin layer of gold was coated on the sample.

**Washing/etching/removing gold from the molecular crystal and substrate:**

Lugol's Iodine solution was used to etch the gold from the substrate.

<u>Preparation of Lugol's Iodine solution:</u> 10g of potassium iodide was dissolved in 30 ml of distilled water. 5 g of iodine was added to it and heated gently with constant stirring until it was dissolved. Later, the solution was diluted to 100 ml with distilled water.

<u>Etching:</u> The FIB milled micro-sheets containing coverslip was placed in a petridish. The freshly prepared Lugol's iodine solution was added gently via the walls of the petridish. The coverslip was allowed to soak for about 2 minutes, then the solution was pipetted out using a dropper. Later, it was washed with distilled water. This procedure was repeated again with a soaking time of 2 minutes (figures 5.a-f).

**Polarised light microscope:**

The images of fabricated and milled micro-sheets were captured using NIKON eclipse LV100N POL polarising microscope. It was equipped with an epi-illuminator (NIKON 12V 50W), DS-Fi3 camera having 5.9 megapixel CMOS sensor, which enables superior color reproduction and NIKON TU plan fluor EPI P series objectives (4x, 10x, 20x and 50x) for pin-sharp aberration free images regardless of magnification.

**Confocal micro-spectroscope:**

Fluorescence spectra of the micro-sheets were recorded on a WI-Tec confocal spectrometer equipped with a Peltier-cooled CCD detector. Using 300 grooves/mm grating BLZ = 750 nm. All measurements were performed in transmission mode geometry. A solid state 405 nm laser was used as an excitation source. To collect the output signals from the specific area of micro-sheets, 150x objective (N. A.: 0.95) was used. For acquiring single spectrum before FIB milling, the laser power, integration time and accumulations were optimised to 0.05 mW, 2 s and 5, respectively. Whereas, after FIB milling, it is 4 mW power, 1 s integration time and 20 accumulations for acquiring a single spectrum. To capture the dark field images higher laser powers were used (before milling: 0.1 mW, after milling: 15 mW). The images were processed by using WI-Tec 5.2 software.

**Fluorescence Lifetime Imaging:**

FL decays and FL lifetime images were recorded on a time-resolved (Micro-Time 200, Pico Quant) confocal FLIM setup, which was equipped with an inverted microscope (Olympus IX 71). Measurements were performed at room temperature, on a micro-sheets deposited cover-slip. The samples were excited by a 405 nm ps diode pulse laser (power ≈ 5 μw) with a stable repetition rate of 20 MHz (FWHM: 176 ps) through a water immersion objective (Olympus UPlans Apo; 60 ×; NA 1.2). Signal from the samples was collected by the same objective and passed through the dichroic mirror, filtered by using a 430 nm long-pass filter to cut off any exciting light. The signal was then focused onto a 50 μm diameter pinhole to remove the out-of-focus signal, recollimated, and directed onto a (50/50) beam splitter prior to entering two single-photon avalanche photodiodes. The data acquisition was carried out with a SymPhoTime software-controlled PicoHarp 300 time-correlated single-photon counting module in a time-tagged time-resolved mode. The overall resolution of the setup was 4 ps.

**FESEM analysis:**

A thin layer of gold was coated on the substrate using 15 μA current for 80 sec. The size and morphology of the milled micro-sheets were examined by using a Zeiss field emission scanning electron microscope operating at an accelarating voltage of 5 kV.